\documentstyle[11pt,aaspp4]{article}

\slugcomment{ApJ Letter, accepted, Mar. 5, 2000}

\lefthead{Bobing Wu \& Edward Fenimore}
\righthead{Gamma-Ray burst Spectral lags}
\begin{document}
\title{Spectral Lags of Gamma-Ray Bursts From Ginga and BATSE}
\author{Bobing Wu$^{1,2}$ and Edward Fenimore$^{1}$}
\affil{$^1$MS D436, Los Alamos National Laboratory, Los Alamos, NM
87545}
\affil{$^2$LCRHEA, Institute of High Energy Physics, Beijing 100039,
China}
\begin{abstract}
The analysis of spectral lag between energy bands,
which combines temporal and spectral analyses,
can add strict constraints to gamma-ray burst (GRB)
models. In previous studies, the lag analysis
focused on the lags between channel 1 (25-57 keV)
and channel 3 (115-320 keV) from the Burst and
Transient Source Experiment (BATSE).
In this paper, we analyzed the cross-correlation
average lags (including approximate uncertainties)
between energy bands  for two GRB samples: 19
events detected by Ginga, and 109 events detected
by BATSE. We paid special attention to the BATSE
GRBs with known redshifts, because there has been
a reported connection between lag and luminosity.
This extends our knowledge of spectral lags to
lower energy ($\sim 2$ keV). We found that lags
between energy bands are small. The lag between
the peak of $\sim 50$ keV photons and $\sim 200$ keV
photons is $\sim 0.08$ sec. The upper limit in the
lag between $\sim 9$ keV photons and $\sim 90$ keV
photons is $\sim 0.5$ sec. Thus, there are not
large shifts at low energy. We found that about
20\% of GRBs have detectable lags between energy
bands in the Ginga and BATSE samples. From the internal
shock model there are three sources of time structure
in GRB pulses: cooling, hydrodynamics, and angular
effects. We argue that cooling is much too fast to
account for our observed lags and angular effects
are energy independent. Thus, only hydrodynamics
can produce these lags. Perhaps the radiation process
varies as the reverse shock moves through the shell.

\end{abstract}

\keywords{gamma-rays: bursts}

\newcount\eqnumber
\eqnumber=1
\def\neweq{{\the\eqnumber}\global\advance\eqnumber by 1}
\def\eqnam#1#2{\xdef#1{\the\eqnumber}}
\def\lasteq{\advance\eqnumber by -1 {\the\eqnumber}\advance
    \eqnumber by 1}
\def\domega{{\rm d}$\Omega$}
\def\Mesz{M\'esz\'aros}
\def\INITIALCON{{4\pi E_{54} \over \rho_0{\rm d}\Omega}}
\section{INTRODUCTION}

More than 26 years have passed since gamma-ray
bursts (GRBs) were discovered (\cite{Klebesadel73}).
The Burst and Transient Source Experiment (BATSE)
on the {\it Compton Gamma Ray Observatory (CGRO)}
found that GRBs appear to be isotropic on the sky,
yet there is a dearth of faint events compared to
the brightest events, implying that the bursts are
at cosmological distances (\cite{Meegan92}).  The
cosmological origin of many GRBs was firmly
established as a result of follow-up observations
of fading X-ray counterparts to GRB sources
discovered with the Beppo-SAX mission
(e.g., \cite{Costa97}). However, the radiation
mechanism of GRBs is still unclear. This situation is
partly due to the absence of any significant
correlations between the GRB spectral and
temporal properties.

Recently, several research groups have investigated
the correlation between GRB spectral and temporal
properties. By using the average autocorrelation
function and the average pulse width,
Fenimore et al. (1995) presented a very well defined
relationship: the average pulse width of many
bursts, $\Delta \tau$, is well fitted by a power
law of the energy $E$ at which the observation is
made: $\Delta\tau \simeq 18.1~E^{-0.45 \pm 0.05}$.
Norris et al. (1996) proposed a ``pulse paradigm''
and also found that the average pulse shape
dependence on energy is approximately
a power law, with an index of $\sim -0.40$,
consistent with the  analysis of
Fenimore et al. (1995). Kazanas, Titarchuk \& Hua (1998)
proposed that synchrotron cooling could account
for the power law relationship between
$\Delta \tau$ and $E$.

The general observed trend in spectral evolution is
hard to soft (\cite{Norris86}, \cite{Norris96},
\cite{Band97}). The hard-to-soft spectral evolution
can lead to a distinct, observed effects: pulse
peaks migrate to later times and become wider at lower
energies (\cite{Norris96}, \cite{Norris99}).
Cheng et al. (1995) claimed that about 24\% of bursts in
their sample of BATSE bursts have detectable
time delay between BATSE channel 1 (25-57 keV) and
channel 3 (115-320 keV). In this paper, we analyzed
the cross-correlation average lags between different
set of energy bands  for two GRB samples: 19 events
detected by Ginga, and 109 events detected by BATSE.
We discussed our results and its implication
for GRB models.

\section{METHODOLOGY}

The cross-correlation function (CCF) has been widely
used to measure the temporal correlation of two
GRB energy bands $c_{1i}$ and $c_{2i}$ (\cite{Link93},
\cite{Cheng95}, \cite{Norris99}). Here $c_i=m_i-b_i$
is the net counts from GRB time profiles, where the
background contribution, $b_i$, has been subtracted
from raw counts $m_i$. A time interval of T is
selected around the largest peak consisting of
$N$ time bins each of duration $\Delta T$ s,
indexed between $-N/2$ and $+N/2$. The CCF as a
function of time lag, $j \Delta T$, is

\begin{eqnarray*}
CCF_j & = & {\sum\limits_{i=-N/2}^{N/2}{{c_{1i+j} c_{2i}} \over S}}
~~~~~~~~~{\rm for} ~~~~j\not = 0 \\
& = & 1 ~~~~~~~~~~~~~~~~~~~~~~~~~~~{\rm for}~~~~j = 0
\end{eqnarray*}

\noindent where the normalization factor $S$ is

$$ S= {\sum\limits_{i=-N/2}^{N/2} (c_{1i} c_{2i}-\sqrt{m_{1i} m_{2i}})}
$$
The $\sqrt{m_{1i} m_{2i}}$ term in S
normalizes the CCF so that coherent noise at $j=0$ is
accounted for.

Here, we must point out that it is not always
reliable to find the lags by the CCF,
especially when the observed profiles are
relatively smooth or these is strong spectral
evolution.  The reason, in part, is that
the CCF is an average quantitative description
of more than one peak in two time series.
Norris et al. (1996) analyzed individual peaks by
obtaining best fit positions, intensities, and
widths for each energy channel.  From this, we can
tell if the CCF lag, indeed, tells us the
lag between two energy bands.  A worst case might
be BATSE burst 1085. Burst 1085 was fitted with 4
pulses for channel 1 and channel 3. The peak shifts
for the pulses are 0.657 s, 1.128 s, 1.181 s
and $-0.160$ s (\cite{Norris96}).  However,
the CCF of the observed profile yields a time
shift of 2.624 s between channel 1 and channel 3
(\cite{Cheng95}).  Another example is BATSE burst
543, which was fitted with 4 pulses for channel 1
and channel 3, and the peak shifts for the
pulses are $-0.002$, $-0.015$, 0.06 and 0.09 s
(\cite{Norris96}), while the CCF of the observed
profile yields a time shift of 0.256 s between
channel 1 and channel 3 (\cite{Cheng95}).
In cases where the lags are small the CCF lag
and the pulse fit lag usually agree.

Thus, pulse fitting and CCF can give very different
results. One must be careful to use them appropriately.
To use the CCF lag, one must always compare to a
model which has calculated two energy ranges and
a resulting expected CCF. The pulse fitting could
provide a more direct measurement of a lag
than the CCF. Unfortunately, we do not often have
statistically significant samples in most of the
peaks, especially in Ginga.

We will select bright GRB events from two samples:
the 4th BATSE catalog and Ginga. From the current
4th BATSE catalog (\cite{paciesas99}),  we use the
4-channel LAD DISCSC data with a time resolution of
64 ms which meet the criteria of $T_{90}$ $>$ 2s
and fluence $>$ 5 photon cm$^{-2}$ s$^{-1}$.
We set the criteria for two main reasons:
1) short bursts are not suitable for timing analysis
with a time resolution of 64 ms; 2) the strong bursts
provide good statistics. This resulted in a total
of 109 usable events.

The Ginga GBD (Gamma-ray Burst Detector) was in
operation from March 1987 to October 1991.
During this time $\sim120$ GRBs were identified
(\cite{Ogasaka91}, \cite{Fenimore93}).
The GRB detectors on Ginga  consisted of a
proportional counter (PC) sensitive to photons
in the 2-to-25 keV range and a scintillation
counter (SC) recording photons with energies
between 15 and 350 keV. The temporal resolution
of the time history data depended on the telemetry
mode.  The on-board trigger system was very similar
to BATSE. It checked the 50-to-380 keV count rate
for a significant increase ($11\sigma$ on
either 1/8, 1/4, or 1 s time scales). Upon such a trigger,
special high resolution temporal data (called
``Memory Read Out'' [MRO]) would be produced.  The MRO
time history has a time resolution of 0.03125 s.
The PC MRO time history extends from 32 s before the
trigger to 96 s after the trigger.  The SC MRO time
history extends from 16 s before the trigger to 48 s
after the trigger. Besides the time history data,
the MRO also contains spectral data recorded
at 0.5 s intervals.

We selected from the 120 Ginga GRBs a sample of 19 MRO
events for which statistically good light curves were
available, for which the events were entirely covered
by the MRO interval, and for which we could be reasonably
certain that the burst occurred within the forward,
$\pi$ steradian field of view of the detectors
(front-side events.)

We selected three sets of low and upper energy
ranges to calculate the spectral lags:
$CCF_{x\gamma}$, $CCF_{pcsc}$ and $CCF_{13}$. From
the Ginga sample, $CCF_{x\gamma}$ is based
on 2-10 keV and 50-100 keV count rates from the
MRO spectral data (0.5 s resolution). Also
from Ginga, $CCF_{pcsc}$ is based on the PC
count rates (2-25 keV) and the SC count rates
(15-350 keV) from the MRO time histories
(31.25 ms resolution). The BATSE $CCF_{13}$ is
based on channel 1 (25-57 keV) and channel 3
(115-320 keV) and has 64 ms resolution.

Before computing the CCF, the background
must be estimated and subtracted from the observed
profiles to yield signal profiles. For majority of
the analyzed bursts, a linear fit or quadratic fit
was reasonable and it was unnecessary to propagate
the background uncertainty into the CCF.

If the observed data had much finer resolution than
GRB temporal features and high signal-to-noise
levels, the CCF curves would be very smooth and
its side lobes would be much lower than its central
peak. We could find the lag by simply recording the
lag of the CCF peak itself (e.g., as was done by
\cite{Cheng95}). In our actual Ginga samples, the
time resolution and/or SNR are not good enough.
Norris et al. (1999) measured the lag in BATSE
bursts by fitting the peak of the CCF and finding
the peak location from the best fit function.
Neither methods provide an uncertainty of the lag.
One goal of this paper is to obtain approximate
uncertainties via a bootstrap method
so we can determine the significance of the lags.

\section{AVERAGE LAGS WITH APPROXIMATE UNCERTAINTIES}

To obtain spectral lags with approximate uncertainties
 \footnote{The uncertainties provide a measure of the
 stability of the calculation of the $\tau$  estimates.},
 we did  10,000 Monte Carlo realizations from the observed
 time histories for each burst, and therefore got 10,000
 lags for each burst by recording the lag of the
 CCF peak. From this method, we could find the mean lag
 and its approximate uncertainty (i.e., the region that
includes $68.3\%$ of the realizations) for
each burst and the average lag from all GRBs for
each set of energy ranges.

For each realization, we selected a count sample for each
time bin and each spectral bin from a poissonian
distribution based on the observed gross counts
(including background). We then removed the background
and calculated the CCF. Based on the Monte Carlo
realizations for each burst, we computed the
average lag and its variance for each set of
energy bands.

To compute the CCFs, a time interval
$T=N \Delta T$ of 32 s was used
for the Ginga bursts, and $T=N \Delta T$ of
32.768 s for the BATSE bursts. If the duration
of a burst is less than the time interval, the whole
profile of the burst was used.

Table 1 summarizes our results for two sets of
energy bands for the Ginga bursts.
The main difference between them is that
$\tau_{pcsc}$ is based on time history MRO
data (combining two samples together to give
62.5 ms resolution), and $\tau_{x\gamma}$ uses
the spectral MRO data (0.5 sec resolution).
The detailed results for BATSE GRBs are not
listed in the paper. In Table 1, the burst date is
given by year/month/day in the first column,
the next two columns ($\tau_{x\gamma}$ and
$\sigma _{x\gamma}$) are  the average lag
and its standard deviation for energy ranges
x (2-10 keV) and $\gamma$ (50-100 keV). The next
two columns ($\tau_{pcsc}$ and $\sigma _{pcsc}$)
are  the average lag and its standard deviation
for energy ranges PC (2-25 keV) and SC (15-350 keV).
In the last two columns we list the duration ($T_{90}$)
and peak intensity (50-300 keV) based on the
SC time histories.

Based on the $CCF_{x\gamma}$ from Ginga, 6 out
of 19 GRBs ($\sim32\%$) have detectable lags at
a $1\sigma$ level (870707, 880205, 880725, 890929,
901001, and 910815); for $CCF_{pcsc}$, 5 of 19
GRBs ($\sim26\%$) have the lags with $1\sigma$ (880205,
880725, 890929, 900126, and 910206). The difference between
$CCF_{x\gamma}$ and $CCF_{pcsc}$ is not significant
and is caused by the different spectral ranges and
the different time resolution.

In our BATSE sample, (a): 36 out of 109 GRBs
($\sim33\%$) have lags less than $0.0+1\sigma$ s;
(b): 53 out of 109 GRBs ($\sim49\%$) have lags
less than $0.064+1\sigma s$; and (c): 20 of 109
GRBs ($\sim18\%$) have lags greater than
$0.064+1\sigma$ s.
(Here, we have compared  the lags against a
rough combination of the quantization
uncertainty [64 ms] and the statistical
uncertainty.) The lags for class (a)
are not detectable within the current
time resolution; the reality of lags for class (b)
are questionable because the lags are very close
to the uncertainty;  and the lags for class (c)
are fairly reliable lags. The fraction in our sample
with lags is roughly consistent with the results of
Cheng et al. (1995), who claimed that about $24\%$
of the bursts in their sample of BATSE bursts have
detectable time delay. However, Cheng et al. (1995)
included a few bursts in that 25\% whose lags were
0.064 s, equal to the time resolution.
(Remember, Cheng et al. (1995) identified the
value of the lag as which 64 ms CCF sample was
the largest, so the lags had to be multiples of 64 ms.)
Our criterion is more strict, requiring a lag to be
at least bigger than the time resolution plus 1$\sigma$.
Even comparing to $0.064+1\sigma$ s might count some
events as having lags when, in fact, the lag arose
from statistical variations. Thus, we probably
actually found more events with lags than did
Cheng et al. (1995). Also, about 10\% in the
sample  have lags more than $3\sigma$ from zero,
and only 4\% have  negative lags.

Figure 1 is the distribution of lags from the Monte
Carlo realizations. There are 190,000 and 1,090,000
entries for Ginga and BATSE, respectively. These
distributions are not estimates of the distribution
of the true lags in the samples as might be found
from pulse fitting. Rather, the distributions are
very roughly a convolution of the true lag distribution
with a function describing the measurement accuracy
of using the CCF to estimate the lags. These distributions
should only be compared to models where the lags are
determined by a CCF of two energy ranges from the model.
Figure 1a is the distribution of CCF lags $\tau _{x\gamma}$
for 19 GRBs from Ginga with a temporal resolution
of 0.5 s. The mean lag for the sample is
$\langle \tau _{x\gamma} \rangle = 0.32_{-0.48}^{+0.44}$ s.
Figure 1b is the distribution of lags $\tau _{pcsc}$
for 19 GRBs of Ginga with the temporal resolution
of 0.0625 s. The mean lag for the sample is
$\langle \tau _{pcsc} \rangle = 0.19_{-0.26}^{+0.32}$ s.
Figure 1c is the distribution of lags $\tau _{13}$
between channel 1 and channel 3 for 109 GRBs of BATSE
sample with the temporal resolution of 0.064 s.
The mean lag for the sample is $\langle \tau _{13}
\rangle = 0.077_{-0.072}^{+0.034}$ s.

We used an average energy to qualitatively estimate
the energy range represented by the CCF lag. We
assumed a power law spectra with index of $-1.5$
and use typical response functions to estimate the
average photon energy in each energy range. The
average energies for BATSE's channel 1
(25-57 keV) \& channel 3 (115-320 keV) are 48 keV and
193 keV, respectively. The average lag between 48 keV
and 193 keV is less than $0.08$ s. The average energies
for Ginga's PC (2-25 keV) \& SC (15-350 keV) are 9 keV
and 86 keV, respectively. The average lag between 9 keV
and 86 keV is less than $\sim$0.5 s. The above qualitative
results show that the peak of the emission is not delayed
substantially at lower energy.

We note that the uncertainties of the (average) lags
from Ginga are relatively large, which is mainly
due to the limited signal-to-noise ratio (SNR).
The BATSE events have much higher SNR than Ginga's,
therefore the uncertainty of the lags for BATSE
sample is relatively small.

Norris et al. (1999) has used lags from
cross-correlations of BATSE data as a predictor
for absolute luminosity.  Based on six GRBs with
known red-shifts, they found that the (isotropic)
luminosity is approximately $1.3 \times
({\tau_{13}/0.01 {\rm s}})^{-1.14} \times 10^{53}$ erg s$^{-1}$.
We calculated the lag somewhat differently than
Norris et al.(1999). They included data above
a fixed fraction of the highest peak and interpolate the
cross-correlation function with a quadratic equation.
We used the whole time history and performed Monte Carlo
realizations to determine the uncertainty for the lag.
In Table 2 we have analyzed four of the six BATSE GRBs
with redshifts. (Only four had publicly available time
histories.)  The first column is the burst date,
the second and third columns are the luminosity and
$\tau_{13}$ lag as reported by Norris et al. (1999),
the fourth column is our lag, the fifth column is
our 1$\sigma$  uncertainty, the sixth column is our
average lag found with a quadratic interpolation of
each realization, and the last column is the
1$\sigma$ uncertainty found the quadratic
interpolation of each realization. Note that we tend
to find larger lags than Norris et al (1999)
by a factor of 3 to 7. The difference is not
due to the use of quadratic interpolation because that
technique gives values very close to the average
of many realizations (see Table 2). Perhaps the difference
is due to the fact that Norris et al. (1999) takes only
the tips of peaks and we use the whole time history
including times when the signal is at background.
In any case, it is clear that the luminosity-lag
relationship is strongly dependent on how the lag
is defined.

\section{DISCUSSION}

In the standard fireball scenario, the most likely
radiation process in GRBs is synchrotron emission
(\cite{Katz94}; \cite{sari96}). Synchrotron emission
often gives a spectral-temporal correlation
$t_{\rm syn}(E)~\propto~E^{-0.5} $  which is not very
different from the observed correlation
$\Delta\tau \propto~E^{-0.45 \pm 0.05}$  (\cite{Fenimore95}).
Kazanas, Titarchuk and Hua (1998) claimed that
the Fenimore et al. $\Delta\tau$ relationship rises
from synchrotron cooling. In the synchrotron cooling
model, the electrons  cool, and the electron's average
energy becomes smaller, which causes the emission
peaks at lower energy at later time. But, if the lags
we found in this paper are caused by the cooling
process, the typical magnetic field should be $\sim 100$ gauss.
In fact, most of the current GRB models require a much
strong magnetic field (e.g., \cite{Piran99}) which
leads to very fast cooling. For example, by using typical
parameters in internal shock model, the observed cooling
time $\tau_{cool}$ at a given frequency is (\cite{Piran99}):
$$
\tau_{cool} (h \nu)~\sim~2 \times 10^{-6}~{\rm sec}~ \epsilon_B^{-3/4}
\left({h\nu_{obs}\over100~{\rm keV}}\right)^{-1/2}
$$
where the dimensionless equipartition parameter $\epsilon_B$
is the ratio of the magnetic field energy density to the
total thermal energy $e$: $\epsilon_B~={B^2\over8\pi e}~$,
and its typical value is $0.01$. Thus, the typical cooling
timescale is $6.3 \times 10^{-5}$ sec. We conclude that the
similarity noted by Kazanas, Titarchuk and Hua (1998) between
synchrotron cooling and the Fenimore et al. (1995) pulse
width-energy relationship is actually a coincidence.
The synchrotron cooling cannot simultaneously explain the large
flux without a large magnetic field and the delays without a
smaller magnetic field.

What then could produce the delays? There are three contributers
to the pulse width in the external shock model of GRB pulses:
cooling, hydrodynamics, and angular spreading. To produce the
observed flux, the strong magnetic field produces a
cooling time that is much shorter than
the hydrodynamic time scale and the angular time scale.
The hydrodynamic time scale is the time it
takes for the reverse shock to cross the shell,
$\tau_{\rm hydro} = 2\Gamma_{rs}^2 \Delta T$
(\cite{Kobayashi97}). The angular time scale is
the time it takes for off axis photons to arrive
at the detector,
$\tau_{\rm ang} = 2\Gamma_{\rm new}^2 \Delta T$
where $\Gamma_{\rm new}$ is the Lorentz factor after
the two shells collide. The resulting time profile
is a convolution of the three processes. The cooling
time is much shorter while the hydrodynamic and
angular time scales are comparable. Thus, the
time structure (e.g., the lags) in the profile
can only come from the hydrodynamic time (which
dominates the rise of the pulse) and the angular
time scale (which dominates the fall of the pulse).
The angular time arises from kinematics
so has no dependency on energy. Therefore,
the energy dependent lags that we report
here must come from variations associated with
hydrodynamic processes, such as variations in
emission as the reverse shock moves through the shell.
Perhaps density or magnetic field variations cause
the differences we observe.

\acknowledgments
We thank the referee for very extensive and detailed
report on this manuscript.
This work was done under the auspices of the US Department
of Energy.

%




\clearpage

\figcaption[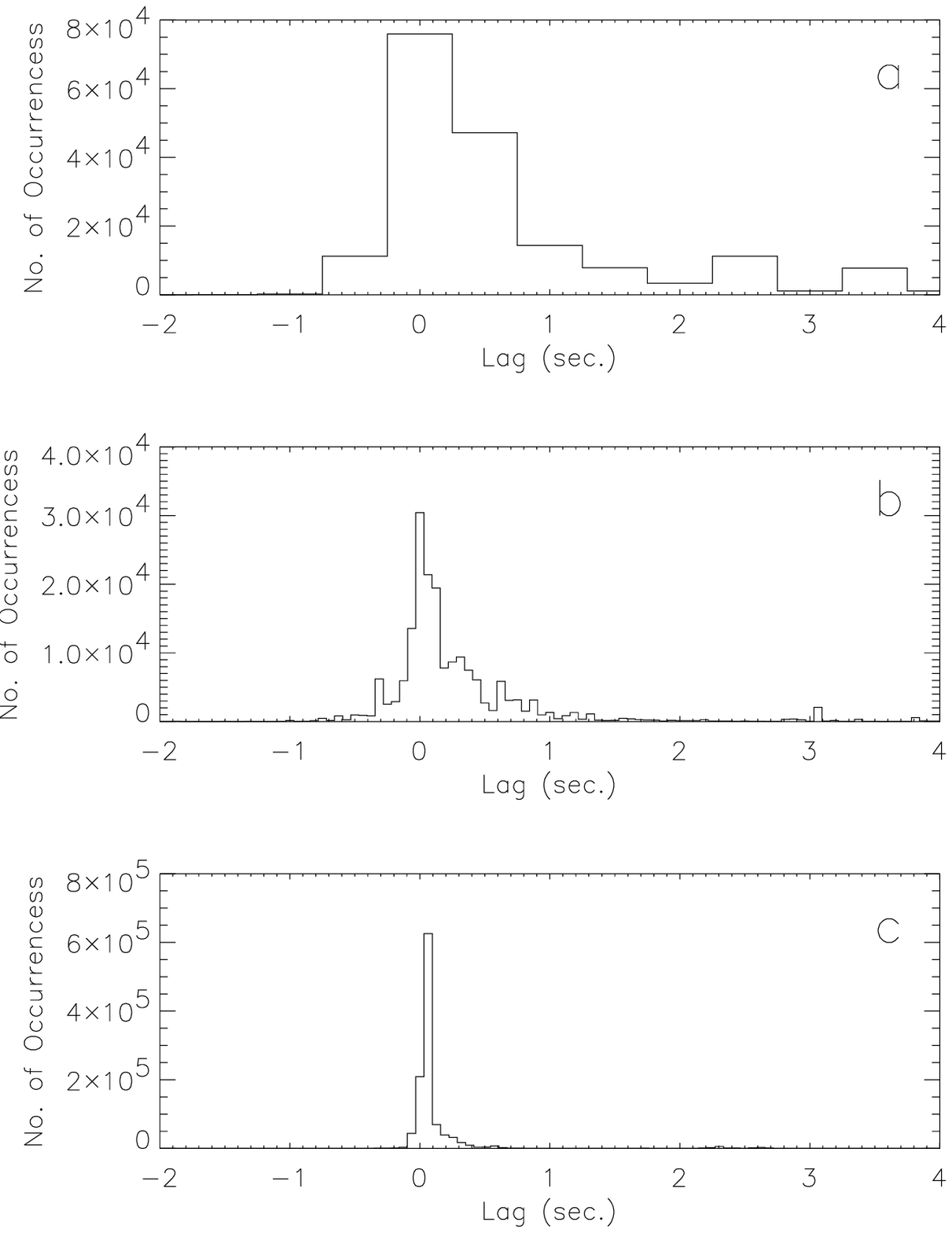]{
Distributions of lags from Monte
Carlo realizations. For each bursts, 10,000 realizations were
done. (a): Distribution of lags $\tau _{x\gamma}$ from the  CCF
of the Ginga x (2-10 keV) and $\gamma$ (50-100 keV) spectral
histories for 19 GRBs
with the temporal resolution of 0.5 s. Based on the 190,000
realizations, $\langle \tau _{x\gamma}
\rangle = 0.32_{-0.48}^{+0.44}$ s. (b): Distribution
of lags $\tau _{pcsc}$ from the  CCF of the Ginga
PC (2-25 keV) and SC (50-100 keV) time histories for 19 GRBs
with the temporal resolution of 0.0625 s.
Based on the 190,000 realizations, $\langle \tau _{pcsc}
\rangle = 0.19_{-0.26}^{+0.32}$ s. (c): Distribution
of lags $\tau _{13}$ from the CCF of BATSE
channel 1 (25-57 keV) and channel 3 (115-320 keV) for 109 GRBs
with a temporal resolution of 0.064 s.
Based on the 1,090,000 realizations, $\langle \tau _{13}
\rangle = 0.077_{-0.072}^{+0.034}$ s. Above results
show that the peak of the emission is not delayed
substantially at lower energy.
 \label{fig1}}

\clearpage
\begin{deluxetable}{crrrrcc}
\tablecaption{Mean lags for Ginga sample from Monte Carlo Realizations.
 \label{tbl-1}}
\tablewidth{0pc}
\tablehead{
\colhead{Burst} &
\colhead{$\tau_{x\gamma}$}   &
\colhead{$\sigma_{x\gamma}$}  &
\colhead{$\tau_{pcsc}$} &
\colhead{$\sigma_{pcsc}$}  &
\colhead{duration} &
\colhead{peak intensity} \\
\colhead{} &
\colhead{(s)} &
\colhead{(s)} &
\colhead{(s)} &
\colhead{(s)} &
\colhead{(s)} &
\colhead{$ \rm (photons/cm^2/s)$}
}
\startdata

   870303  &   1.921  &  2.255 &  0.284  & 0.321  & 60.00  & 1.30 \nl
    870707  &   1.521  &  1.103 &  0.050  & 0.138 & 7.00   & 3.66 \nl
    880205  &   3.459  &  0.536 &  4.210  & 0.887 & 45.00   & 5.83 \nl
    880725  &   0.367  &  0.332 &  0.247  & 0.166 & 11.00  &  3.91 \nl
    880830  &   0.216  &  0.535 &  0.636  & 0.902 & 25.00  &  1.50 \nl
    881117  &  $-$0.083  &  1.500 &  0.050  & 0.317  & 1.50 & 4.30 \nl
    890330  &  $-$1.568  &  5.377 &  0.146  & 3.274 & 20.00  & 4.06 \nl
    890929  &   1.059  &  0.497 &  0.782  & 0.371 &  2.00 &   3.33 \nl
    900126  &   0.000  &  0.075 &  0.158  & 0.056 & 13.00   & 4.12 \nl
    900129  &   0.410  &  0.439 &  0.041  & 0.181 & 10.00  & 3.61 \nl
    900623  &  $-$0.292  &  0.418 & $-$0.049  & 0.263 & 14.00  & 2.04
\nl
    900919  &   0.000  &  0.080 &  0.000  & 0.010 & 0.25  & 4.16 \nl
    900928  &   0.356  &  0.428 &  0.049  & 0.163 & 4.00 &  4.57 \nl
    901001  &   0.508  &  0.506 &  0.120  & 0.437 & 20.00  & 2.84 \nl
    901009  &   0.622  &  0.748 &  0.427  & 0.602 & 16.00  & 2.75 \nl
    910206  &   1.449  &  2.221 &  0.345  & 0.198 & 13.00  &  3.38  \nl
    910402  &  $-$0.016  &  0.089 &  0.008  & 0.112 & 48.00  & 4.96 \nl
    910815  &   0.565  &  0.530 &  0.435  & 0.543 & 35.00  &  0.96 \nl
    910821  &   0.271  &  0.418 &  0.170  & 0.602 & 80.00  &  0.84 \nl

\enddata

\end{deluxetable}

\clearpage
\begin{deluxetable}{crrrrrr}
\tablecaption{Spectral lags for GRBs with red-shifts.
 \label{tbl-2}}
\tablewidth{0pc}
\tablehead{
\colhead{GRB Date} &
\colhead{$\rm Luminosity^1$}    &
\colhead{$\tau_{13}^2$}   &
\colhead{$\tau_{13}^3$}   &
\colhead{$\sigma_{13}^4$}  &
\colhead{$\tau_{13}^5$}    &
\colhead{$\sigma_{13}^6$}
}
\startdata
970508.904  & 2.9 &  0.140  &  0.381   &  0.222   &  0.440 &  0.220 \nl
971214.973  & 17.3 &  0.007  &  0.055   &  0.058   &  0.052 &  0.037 \nl

980703.182  & 8.9 &  0.110  &  0.365   &  0.265   &  0.364  & 0.272 \nl
990123.407  & 196. &  0.012  &  0.063   &  0.037   &  0.079 &   0.029
\nl
\tablenotetext{1}{Luminosity from \cite{Norris99}, unit is $10^{51}
ergs~s^{-1}$.}
\tablenotetext{2}{$\tau_{13}$ reported by \cite{Norris99}.}
\tablenotetext{3}{$\tau_{13}$ from  Monte Carlo realizations.}
\tablenotetext{4}{1$\sigma$ uncertainty.}
\tablenotetext{5}{$\tau_{13}$ found with a quadratic interpolation.}
\tablenotetext{6}{1$\sigma$ uncertainty found with a quadratic
interpolation.}
\enddata
\end{deluxetable}



\clearpage

\centerline{Figure 1}
\plotone{lag_st_all.eps}
\clearpage

\end{document}